\documentclass[acmlarge]{acmart}
\usepackage{graphicx}
\usepackage{amsmath}
\usepackage{color}

\graphicspath{ {./media/} }

\usepackage{array}

\AtBeginDocument{%
	\providecommand\BibTeX{{%
			\normalfont B\kern-0.5em{\scshape i\kern-0.25em b}\kern-0.8em\TeX}}}

\acmJournal{DTRAP}

\begin{document}
	
	\title{Scrybe: A Secure Audit Trail for Clinical Trial Data Fusion}
	
		\author{Jon~Oakley}
		\email{joakley@g.clemson.edu}
		\author{Carl~Worley}
		\email{cworley@g.clemson.edu}
		\author{Lu~Yu}
		\email{lyu@g.clemson.edu}
		\author{Richard~R.~Brooks}
		\email{rrb@g.clemson.edu}
		\affiliation{%
			\institution{Clemson University}
			\streetaddress{105 Sikes Hall}
			\city{Clemson}
			\state{South Carolina}
			\country{USA}
			\postcode{29634}
		}
		
		\author{\.{I}lker~\"{O}z\c{c}elik}
		\email{ilker-ozcelik@utc.edu}
		\author{Anthony~Skjellum}
		\email{tony-skjellum@utc.edu}
		\affiliation{%
			\institution{University of Tennessee at Chattanooga}
			\streetaddress{615 McCallie Ave}
			\city{Chattanooga}
			\state{Tennessee}
			\country{USA}
			\postcode{37403}}
		
		\author{Jihad~S.~Obeid}
		\affiliation{%
			\institution{Medical University of South Carolina}
			\streetaddress{171 Ashley Ave}
			\city{Charleston}
			\state{South Carolina}
			\country{USA}
			\postcode{29425}}
		\email{jobeid@musc.edu}
		
		\renewcommand{\shortauthors}{Oakley and Worley, et al.}
	
	\begin{abstract}
		Clinical trials are a multi-billion dollar industry. One of the biggest challenges facing the clinical trial research community is satisfying Part 11 of Title 21 of the Code of Federal Regulations \cite{regulations} and ISO 27789 \cite{iso27789}. These controls provide audit requirements that guarantee the reliability of the data contained in the electronic records. Context-aware smart devices and wearable IoT devices have become increasingly common in clinical trials. Electronic Data Capture (EDC) and Clinical Data Management Systems (CDMS) do not currently address the new challenges introduced using these devices. The healthcare digital threat landscape is continually evolving, and the prevalence of sensor fusion and wearable devices compounds the growing attack surface. We propose Scrybe, a permissioned blockchain, to store proof of clinical trial data provenance. We illustrate how Scrybe addresses each control and the limitations of the Ethereum-based blockchains.  Finally, we provide a proof-of-concept integration with REDCap to show tamper resistance.
	\end{abstract}
	
	
	
	\keywords{Blockchain, Clinical Trials,  REDCap,  Secure Audit, Title 21 CFR Part 11, ISO 27789}
	
	\maketitle
	
	%
	
	\section{Introduction} 
	The SARS-CoV-2 pandemic has brought new attention to the clinical trial process. As the digital threat landscape has evolved, this new attention has made it more lucrative for attackers. Recent ransomware attacks have focused on holding SARS-CoV-2 clinical trial data \cite{covid} hostage, but this exposes a critical weakness in how all clinical trial data is stored. The inherent shortcomings in current infrastructure invite attacks on any research group with promising work.
	
	Part 11 of Title 21 of the Code of Federal Regulations defines the controls for electronic records \cite{regulations} imposed by the Food and Drug Administration (FDA). Similarly, ISO 27789 \cite{iso27789} governs the standards for electronic health records (EHR) and audit trails. These controls provide audit requirements that guarantee the reliability of data contained in the electronic records. Codifications regulate clinical trials that are necessary for understanding pathologies, developing new treatments, and improving health. Researchers must guarantee data and consent form authenticity, integrity, and confidentiality.  Electronic Data Capture (EDC) and Clinical Data Management Systems (CDMS) increase the speed and efficiency of clinical studies \cite{edc}, but pose challenges for securing clinical data. Digital information can be easily changed, forged, and fabricated, raising questions about authenticity and integrity.
	
	
	Recently, smart devices and the Internet of Things (IoT) devices have become more common in clinical trials \cite{schobel2015using, lu2016healthcare, kang2018recent}. The prevalence of these devices presents a unique data fusion security challenge, as it combines context-aware computing, traditional sensor fusion, and the regulations that govern the handling and processing of clinical trial research data.
	
	Mobile devices provide electronic questionnaires in a simple format, data validation, and make data aggregation simple. Other devices, like smartwatches, have been used in recent studies ranging from Parkinson's \cite{lopez2019smartwatch} to atrial arrhythmias \cite{koshy2018smart}. These devices generate a rich set of data that must be processed, stored, and analyzed according to the appropriate provenance and security regulations.
	
	A clinical trial planning phase includes creating a study protocol, which specifies the goals, patient groups (or cohorts), pharmaceuticals, and tests. This plan must be approved by the institution's Internal Review Board (IRB) and then be approved by and registered with a government regulatory agency (FDA in the United States). Once the study has been approved, patients are recruited and must provide consent before enrollment into the clinical trial. During the trial, collected data may include physical exam findings, laboratory test results, research questionnaires, and other research data types. After the trial period, the data is analyzed and published in a manner that preserves confidentiality.
	
	Researchers must maintain a clear audit trail, track data creation, modification, and deletion. All actions need to be recorded along with the time and person responsible. REDCap (Research Electronic Data Capture) is a software toolset and database for electronic collection, management of research, and clinical trial data \cite{redcap}. Since REDCap stores acquired data as records, these are the primitive data that must be secured.
	
\begin{figure}
		\centering
		\includegraphics[scale=0.475]{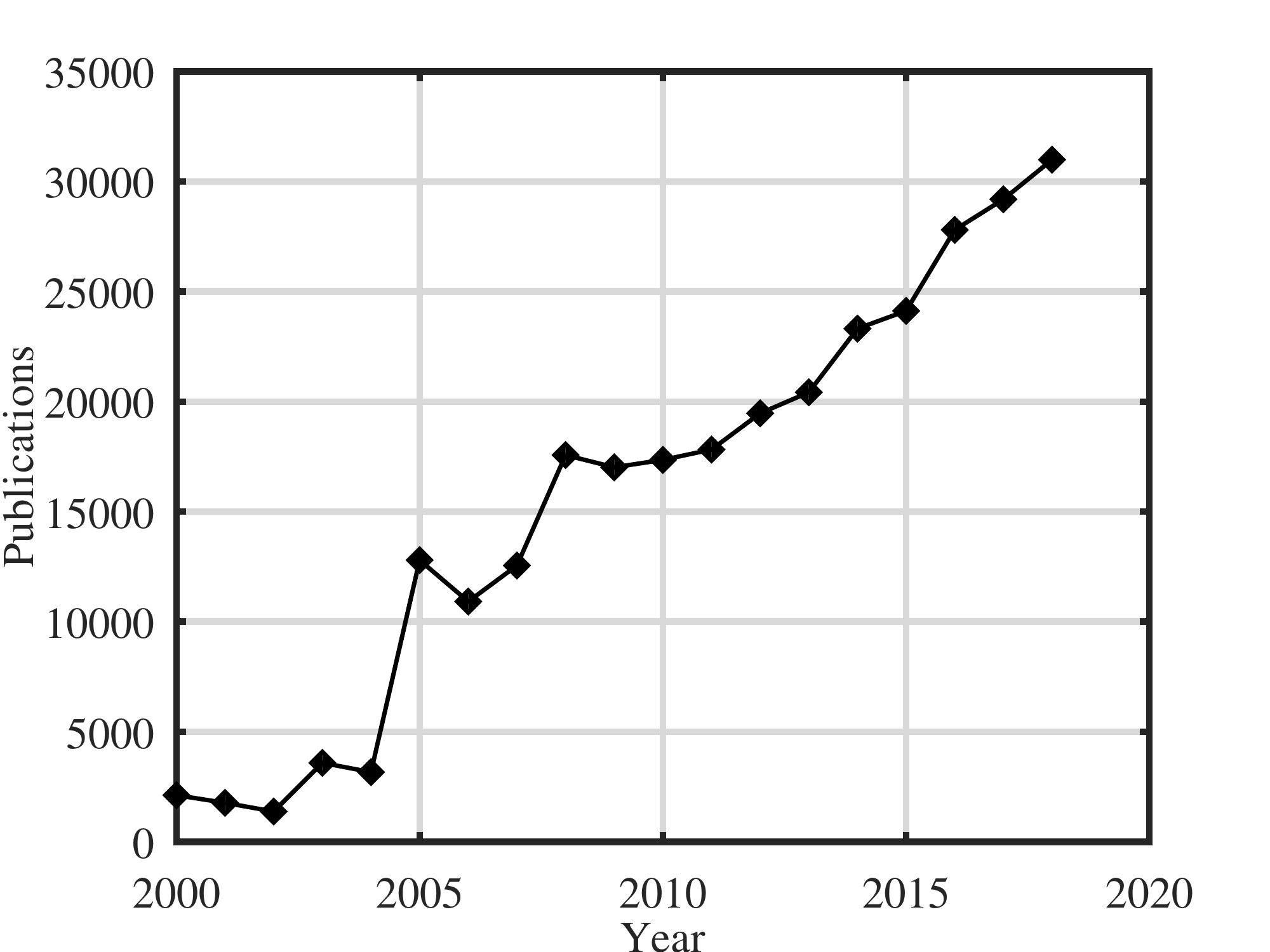}
		\caption{Studies reported to the FDA and registered on ClinicalTrials.gov \cite{num_studies}.}
		\label{fig:study-line}
	\end{figure}

	The number of new clinical trials is increasing each year, compounding the economic impact of securing clinical trial data. In 2018 alone, 30,988 new clinical trials were registered with the FDA \cite{num_studies}. Figure~\ref{fig:study-line} shows the increase in clinical trials over the past decade. As the number of clinical trials grows, so does the amount of data that needs to be secure.
	
	Clinical trial data, and the associated data provenance, is be subject to a large number of external threats \cite{prov_modeling}, but insider threats are also important considerations. A meta-analysis in Science \cite{open2015estimating} was only able to reproduce 39\% of published psychological studies. In a survey of biostatisticians, 31\% of the statisticians surveyed, all active in medical research, reported being involved in a project that knowingly committed academic fraud \cite{ranstam2000fraud}. Scrybe provides an immutable audit trail to satisfy Part 11 of Title 21 and provide controls against insider threat and intentional fraud.
	
	
	Table~\ref{tab:cfr} outlines the controls and governances required for all clinical trial electronic records. From Table~\ref{tab:cfr}, several general categories of controls can be inferred.
	
	\begin{table*}
		\footnotesize
		\centering
		\caption{Controls outlined in Part 11.10 of Title 21, Code of Federal Regulations, for a closed clinical trial electronic records system$^1$ \cite{regulations}.}
		\label{tab:cfr}
		\begin{tabular}{c|c| >{\centering}m{10cm} | >{\centering}m{2.5cm} | c }
			\cline{2-4}
			&Designation & Description & Property & \\
			\cline{2-4}
			\cline{2-4}
			&11.10(a)  & Validation of systems to ensure accuracy, reliability, consistent intended performance, and the ability to discern invalid or altered records. & Integrity \\ Authentication & \\
			\cline{2-4}
			&11.10(b) & The ability to generate accurate and complete copies of records in both human-readable and electronic form suitable for inspection, review, and copying by the agency. Persons should contact the agency if there are any questions regarding the agency's ability to perform such review and copying of the electronic records. & Availability &\\
			\cline{2-4}
			&11.10(c)  & Protection of records to enable their accurate and ready retrieval throughout the records retention period. & Availability &\\
			\cline{2-4}
			&11.10(d) & Limiting system access to authorized individuals. & Access Control  &\\
			\cline{2-4}
			&11.10(e)  & Use of secure, computer-generated, time-stamped audit trails to independently record the date and time of operator entries and actions that create, modify, or delete electronic records. Record changes shall not obscure previously recorded information. Such audit trail documentation shall be retained for a period at least as long as that required for the subject electronic records and shall be available for agency review and copying. & Integrity \\ Authentication \\ Non-repudiation  & \\ 
			\cline{2-4}
			&11.10(g)  & Use of authority checks to ensure that only authorized individuals can use the system, electronically sign a record, access the operation or computer system input or output device, alter a record, or perform the operation at hand. & Access Control &\\
			\cline{2-4}
			&11.10(h) & Use of device (e.g., terminal) checks to determine, as appropriate, the validity of the source of data input or operational instruction. & Integrity &  \\
			\cline{2-4}
			&11.10.(k1)  & Adequate controls over the distribution of, access to, and use of documentation for system operation and maintenance. & Integrity  &\\
			\cline{2-4}
			&11.10.(k2)  & Revision and change control procedures to maintain an audit trail that documents time-sequenced development and modification of systems documentation. & Integrity & \\
			\cline{2-4}
		\end{tabular}
		{\flushleft \scriptsize $^1$Designations 11.10(f), 11.10(i), and 11.10(j) were omitted because they refer to administrative controls outside the scope of this framework.}
	\end{table*}
	
	\begin{itemize}
		\item \textbf{Integrity:} provenance data cannot be falsified and can only be created by authorized parties.
		\item \textbf{Availability}: the provenance system must be consistently able to receive new data, and that data must be able to be consistently viewed.
		\item \textbf{Efficiency:} the system should have a low overhead.
		\item \textbf{Authentication:} provenance data is correctly identified as having come from the correct source.
		\item \textbf{Non-repudiation:} once provenance data has been created, neither creator nor viewer cannot deny its existence.
		\item \textbf{Access control:} only authorized parties can view provenance data. Compliance with the Health Insurance Portability and Accountability Act (HIPAA) privacy and security rules \cite{HIPAA} makes this particularly relevant.
	\end{itemize}
	
	This paper proposes Scrybe, a blockchain-based secure audit trail, and discusses how it addresses the security requirements of clinical trial data outlined in Part 11 of Title 21. A Scrybe proof-of-concept is integrated with REDCap to show these properties and demonstrate the working system.
	
Section \ref{sec:related} discusses existing solutions in this problem space. 	Section \ref{sec:background} provides background information on data provenance, REDCap, and blockchains.  Section \ref{sec:scrybe} introduces Scrybe and outlines the architecture of our system. Section~\ref{sec:clinicaltrial} describes the proof-of-concept integration with REDCap and shows how this architecture satisfies the requirements set from in \cite{regulations}. Section~\ref{sec:discussion} provides a discussion of our results and directions for future work.

	\section{Related Work} \label{sec:related}
	
	Existing blockchain-based provenance solutions for EHR consist of novel blockchain technology and existing blockchain technology. Approaches that propose novel blockchains often overlook critical security details.  Limitations in current blockchain technology make a healthcare blockchain built on these technologies impractical for widespread adoption. Unfortunately, there is also a non-trivial group of proposed solutions that claim to use blockchain technology but do not provide enough information to assess the viability of the solution \cite{zhao2017lightweight, xia2017medshare,yang2018medshare,yue2016healthcare,burstiq,linn2016blockchain}.
	
	\subsection{Application-Specific Blockchain Technologies}
	
	One approach is to design a consensus algorithm tailored to the application. These approaches generally address proof-of-work (PoW) efficiency concerns, but many fail to include integrity and security considerations. MediBchain \cite{al2017medibchain,liu2019mbpa} is one such solution. The specifics of the blockchain are very high-level, and concrete details are abstracted. Further, other authors have found security issues with their implementation \cite{xu2019healthchain}. BBDS is a PoW-based blockchain designed to facilitate data sharing \cite{xia2017bbds}. Patel designed a blockchain framework for sharing radiology data using a proof-of-stake (PoS) algorithm \cite{patel2019framework}. Lee and Yang designed a flawed blockchain specific to their work (fingernail microscopy). Peterson et al. designed a blockchain that leverages proof-of-interoperability instead of PoW \cite{peterson2016blockchain}. While the approach is novel, based on the algorithm description, it appears that two malicious nodes could control the network.
	
	\subsection{Tested Blockchain Technologies}	
	Many EHR blockchain solutions build on either Bitcoin or Ethereum. Bitcoin-based solutions \cite{irving2016blockchain,mamoshina2018converging,factom2018} leverage numerous integrations with tools like sidechains and data anchors. Ethereum-based solutions (like MedRec \cite{azaria2016medrec,ekblaw2016case} and others \cite{mytis2017notarization,nugent2016improving,benchoufi2017blockchaina,benchoufi2017blockchainb}) use smart contracts to provide impressive functionality that can easily interface with complicated systems. Other solutions (Healthchain \cite{xu2019healthchain}, MedicalChain\footnote{MedicalChain also relies on the Ethereum blockchain.} \cite{medicalchain}, and \cite{juneja2018leveraging}) are based on Hyperledger Fabric -- a permissioned blockchain developed by IBM. Hyperledger Fabric incorporates functionality present in Ethereum but does not require a cost or resource overhead. Both Bitcoin and Ethereum currently use PoW as their consensus algorithm. While this is currently one of the more popular consensus algorithms, it does not scale well, and it is cost-prohibitive \cite{scherer2017performance}. 
	
	While proof-of-concept solutions based on PoW are popular, they are not long-term solutions for securing data provenance. They result in too much overhead and too little throughput \cite{scherer2017performance}. Ethereum has announced a move to a PoS consensus algorithm. While this is a step in the right direction, it still introduces issues, and any application running on the blockchain will have associated costs required to keep running the smart contracts. With the current cryptocurrency volatility, tying a technological solution to any of these blockchains would be a gamble.  
	
	Hyperledger is a promising permissioned blockchain architecture that offers pluggable consensus algorithms. There are several mainstream choices, such as Hyperledger Fabric, Hyperledger Indy, Hyperledger Iroha, and Hyperledger Sawtooth \cite{hyperledger2017}. There are several popular pluggable ordering/consensus mechanism for Hyperledger Fabric: practical Byzantine Fault Tolerance (PBFT) \cite{castro2002practical}, BFT-SMaRt \cite{bessani2014state}, SBFT \cite{gueta2019sbft}, HoneyBadger BFT \cite{miller2016honey}, and Kafka \cite{kreps2011kafka}. Kafka stands out on this list since it is neither crash-tolerant nor fault-tolerant. The other consensus algorithms are all in the family of PBFT algorithms, which suffer from scalability issues (usually greater than 20 nodes) \cite{vukolic2015quest}. SBFT improves on this, but it still suffers from the other shortcoming of PBFT algorithms -- electing a centralized leader. Similarly, HoneyBadger BFT elects a set of nodes at the beginning of the algorithm that makes it centralized \cite{gilad2017algorand}. Hyperledger Indy uses Robust BFT (RBFT), which is in the same PBFT family that suffers from node scalability \cite{vukolic2015quest}. Hyperledger Iroha uses the Sumeragi consensus, which is based on BChain \cite{duan2014bchain}. Sumeragi also suffers from node scalability issues \cite{hyperledger2017}. Finally, Hyperledger Sawtooth uses PoET as the consensus algorithm. PoET is based on a Trusted Execution Environment \cite{sandell2016tee}. While TEEs are a huge advancement in the space of efficient consensus algorithms, vulnerabilities such as Plundervolt \cite{Murdock2019plundervolt} show there are ways to bypass the trusted environment. 
	
	Our EHR blockchain solution leverages Scrybe's lightweight consensus algorithm \cite{scrybeproof}. This consensus algorithm is designed for a permissioned blockchain, so it does not require excessively wasteful computations. Scrybe's complexity is linear as the number of nodes approaches infinity \cite{scrybeproof}, which addresses the scalability concerns many popular consensus algorithms face. While the current embodiment of our design does not leverage Hyperledger Fabric, a future version may be integrated with a pluggable Scrybe consensus algorithm.
		
	

	\section{Background} \label{sec:background}
	Theoretically, provenance provides several different benefits \cite{goble04}: 1) data integrity, 2) audit trail, 3) replication, 4) attribution\footnote{Attribution implies that changes can be attributed to a given individual or event.}, and 5) information\footnote{Information implies that additional data can be derived from the provenance.}. Provenance tools usually rely on \textit{inversion} or \textit{annotation} \cite{Simmhan:2005:SDP:1084805.1084812,Simmhan:iucstr:2005}. Inversion provenance is the set of transformations required to take an empty data set and arrive at the current state. Annotation involves supplying rich metadata about the changes. Provenance is stored \cite{GD07}: 1) tightly coupled, 2) loosely coupled, or 3) uncoupled. Tightly coupled provenance is stored \textit{in} the data. Loosely coupled provenance is stored \textit{with} the data--it is logically separated, usually in a different file on the same system. Uncoupled provenance is stored \textit{remotely}. The Open Provenance Model (OPM) \cite{Morau2008} and W3C PROV \cite{w3c-prov-primer} provide standards for representing provenance.
	
	
	Various tools have arisen to address the provenance challenge \cite{hambolu2016provenance}. IPython \cite{PER-GRA:2007} is an interactive Python interface that allows users to mix code, text, and media. Taverna \cite{citeulike:12321961} is a Java-based workflow tool that exports OPM-compliant models.  VisTrails \cite{VisTrail:Misc} is a Python-based workflow tool. Karma \cite{KarmaMisc} is an uncoupled provenance tool that conforms to the OPM standard in a client/server relationship. Komadu \cite{suriarachchi2015komadu} is a W3C PROV-based tool that captures system-level events. Kepler \cite{Kepler:2016:Misc} is another environment capture tool that focuses on the execution environment. Swift \cite{4278797} is a provenance capture system for parallel processing. Sumatra \cite{Sumatra:2013:Misc} is another Python-based provenance system focusing on numerical simulations and analyses. Provenance Aware Storage System (PASS) \cite{Carata:2014:PP:2602649.2602651} is an environmental capture tool that captures all system metadata at execution time.
	
	\subsection{REDCap} \label{sec:redcap}
	REDCap is a software toolset and workflow methodology for electronic clinical trial data collection and management \cite{redcap}. REDCap provides web-based tools for data entry, aiding correct entry using real-time validation rules with automated data type and range checks at the time of entry. The system allows the research teams to create and design online surveys and allows survey owners to engage respondents using various notification methods. REDCap data dictionaries can be distributed for reuse at multiple institutions. A library of data dictionaries is made available for standard data collection forms and validated collection interfaces \cite{procurement}.
	
	These features make REDCap an ideal clinical trial collection tool. We use REDCap as our data management system, intending to secure the data provenance. The existing REDCap system already has some valuable security features. The underlying database is typically hosted in secure data centers at the host institutions with layers of redundancy, failover capability, backups, and extensive security checks. The system is inherently compliant with the Health Insurance Portability and Accountability Act of 1996 (HIPAA). It has several layers of protection including, user/group account management, "Data Access Groups\footnote{Data Access Groups allow data to be entered by multiple groups in one database with segmented user rights for entered data}," audit trails for all changes, queries, reports, and Secure Sockets Layer (SSL) encryption. In addition to HIPAA, it can be set up to support other regulatory requirements including Title 21 Part 11 of CFR \cite{regulations} and FISMA-compliant \cite{gantz2012fisma} environments as needed.
	
	Users can export data in native format for several statistical packages, including SPSS, SAS, SATA, R, and comma-separated values files. REDCap has an Application Programming Interface (API), which allows interoperability with external tools, plugins, and mobile apps. REDCap provides an interface for database and case report form creation, either online via a web-based designer or offline using a "data dictionary" spreadsheet template that can be uploaded later into REDCap. This generalized method for quickly creating clinical trial infrastructure has led to  REDCap being used in over 3000 institutions \cite{project_redcap}.
	
	REDCap is written in PHP, and it depends on other software, such as MySQL and the underlying HTTP server \cite{redcapsecurity}. The developers note that REDCap's security depends on this underlying infrastructure, which is known to be vulnerable if improperly configured or not maintained \cite{redcapsecurity}. However, even if the underlying infrastructure is updated, REDCap itself still has documented vulnerabilities.  According to the latest REDCap changelog, \cite{redcapchangelog}, version 9.5.0 (released 12/05/2019) fixed an SQL injection vulnerability that would allow "tech-savvy" users to view any sensitive data they wished.  REDCap is an invaluable tool for researchers, but it is not a secure provenance tool. A malicious user could exploit any of these vulnerabilities to falsify REDCap records.
	
	\subsection{Blockchains} \label{sec:blockchains}
	In 2008, Satoshi Nakamoto proposed Bitcoin, the first cryptocurrency  \cite{nakamoto}. Nakamoto's work extended the ideas presented by Haber and Stornetta \cite{haber1990time}, who presented the first cryptographically secured timestamp audit trail. Bitcoin allowed the digital currency to be exchanged between participants without trusting a central authority (e.g., a bank or wire service). Bitcoin has two categories of participants--\textit{miners} and \textit{users}. Each miner maintains a local copy of a decentralized ledger tracking the balances in all accounts, and any \textit{user} transferring bitcoin broadcasts a signed message that causes all miners to update their local ledgers. 
	
	The blockchain is the data structure each miner uses to store all of the transactions making up the local ledger. Transactions are grouped together in \textit{blocks}, and each \textit{block} contains a cryptographic link to the previous block (hence a ``chain'' of blocks). This cryptographic link is a hash--a one-way function that takes an arbitrary blob of data and returns a random value between 0 and $2^{256} - 1$ with uniform probability. Since the hash is a one-way function, it is computationally hard to determine the original input given the random value. Further, each block is cryptographically \textit{signed}--ensuring its authenticity and integrity.
	
	In Bitcoin, creating a valid block requires a computationally expensive proof-of-work (PoW). This process is called mining. All miners try to guess a random value (nonce) that causes the block's hash to have specific properties. The first miner to guess this random value broadcasts the solution. The process begins again, and the next block contains a hash of the latest solution. The group's consensus is the longest blockchain.
	
	An attacker must create a valid chain that is longer than the current consensus chain to change the consensus. The probability of successfully performing this attack decreases exponentially with each block added to the chain\footnote{The exception to this occurs when the attacker controls more than 50\% of all computing power in the network.}. Mining ensures that the blockchain is an \textit{immutable consensus} among all participants, even in the presence of attackers. One caveat is that the PoW mining process is highly inefficient (computationally and economically), limiting the scalability of the PoW-based systems.
	
	While the most popular use-case is currency transactions, blockchains can be generalized to any use-case where a distributed data store requires consensus and immutability. There are many applications and supporting technologies \cite{generalized} following this trend. Blockchains have also been used to store provenance metadata \cite{prov_modeling}. The immutability property provides strong guarantees of data integrity. The requirement that each block is signed and contains the hash of the previous block ensures signatures are interleaved throughout the blockchain, and each miner's signature validates the signatures of every previous block.
	
	\subsection{Hashes}
	In order to formally prove the fundamental properties of the Scrybe Provenance framework, we provide formal definitions for the cryptographic primitives. A SHA 256 hash function is defined in Equation~\ref{eq:hash} for message $M$, and it has the following relevant properties \cite{appel2015verification}:
	
	\begin{equation} \label{eq:hash}
		\begin{split}
			&H_{256}: M^\alpha \rightarrow \left\{0,1\right\}^{256} \\
			&M^\alpha := \left\{0,1\right\}^\alpha \quad	\forall \alpha \in \mathbb{N}_1
		\end{split}
	\end{equation}
	
	\begin{enumerate}
		\item $H_{256}$ is a one-way function.
		\begin{equation} \label{eq:oneway}
			\nexists H_{256}^{-1}: \left\{0,1\right\}^{256} \rightarrow M^\alpha
		\end{equation}
		
		\item $H_{256}$ is a uniform mapping from the domain to the range.
		\begin{equation}\label{eq:uniform}
			\begin{split}
				\textrm{\textbf{prob}}& \left( H_{256} \left( M^\alpha \right) = = H_{256} \left( M^\beta \right) \right) =  \frac{1}{2^{512}} \approx 0\\
				&\forall \alpha,\beta \in \mathbb{N}_1, \quad M^\alpha \neq M^\beta
			\end{split}
		\end{equation}
	\end{enumerate}
	
	For perspective, the current Bitcoin network hash rate\footnote{The network hash rate is the combined hash rate of all miners currently working to solve the next block.} is $8.5 \times 10^{19}$ hashes per second \cite{hashrate}. At the current rate, it would take $3.42 \times 10^{49}$ years before the Bitcoin network found a collision for a given hash. 
	
	\subsection{Digital Signatures}
	
	Public key encryption assumes two keys: $K_{private}$ and $K_{public}$. The private key is known only to the owner, while the public key is known to everyone.  The formal definition for the public key encryption function is provided in Equation~\ref{eq:encryption}.
	
	\begin{equation}\label{eq:encryption}
		\begin{split}
			&\textrm{Enc}:\left\{0,1\right\}^n \rightarrow \left\{0,1\right\}^n \\
			&\textrm{Dec}:\left\{0,1\right\}^n \rightarrow \left\{0,1\right\}^n
		\end{split}
	\end{equation}
	
	The encryption and decryption functions are asymmetric, as shown in Equation~\ref{eq:symmetric}. $C_{public}$ is the ciphertext when the message is encrypted with the public key, and $C_{private}$ is the ciphertext when the message is encrypted with the private key.
	
	\begin{equation}\label{eq:symmetric}
		\begin{split}
			&\textrm{Enc}\left(M,K_{private}\right) = C_{private} \\
			&\textrm{Dec}\left(C_{private},K_{public}\right) = M \\
			&\textrm{Enc}\left(M,K_{public}\right) = C_{public} \\
			&\textrm{Dec}\left(C_{public},K_{private}\right) = M \\
		\end{split}
	\end{equation}
	
	From this definition, we can see that anything encrypted using a private key can be decrypted using the public key. Until recently, RSA has been the standard for public-key cryptography \cite{rivest1978method}. Elliptic curve cryptography leverages different mathematical principles to reduce the overall key size (significantly) at a slight performance cost \cite{hankerson2011elliptic}. Both of these approaches are susceptible to quantum attacks \cite{quantum}, but those are distant concerns.
	
	\subsection{Integrity}\label{sec:integrity}
	
	Scrybe uses cryptographic signatures to ensure the integrity of data provenance. The formal definition for a cryptographic signature, shown in Equation~\ref{eq:signature}. 
	
	\begin{equation}\label{eq:signature}
		\begin{split}
			&S: \left\{0,1\right\}^\alpha \rightarrow \left\{0,1\right\}^{256} \quad \forall \alpha \in \mathbb{N}_1 \\
			&S(M) := \textrm{Enc}\left( H_{256}\left( M \right), K_{private}\right)
		\end{split}
	\end{equation}
	
	The signature, $S(M)$, is appended to the message, $M$, so that anyone can verify the message's integrity. Since the assumption is that only the signatory knows the private key, it is assumed that if the decrypted signature, $H_{256}(M)' $, is the same as the hash of the message, $H_{256}(M)$, then the message originated from the owner of $K_{private}$ and the message has not been modified. The transaction containing the hash of the changelog entry is timestamped and cryptographically signed by the researcher, and the block is timestamped and cryptographically signed by the miner. Together, these layers of cryptographic signatures satisfy the integrity property. Using the digital signature standard as documented by the National Institute of Standards and Technology \cite{barker2009digital} guarantees security and portability.
	
	\section{Scrybe Provenance Framework} \label{sec:scrybe}
	Scrybe is a blockchain-based provenance framework that can be adapted to secure clinical trial metadata. Scrybe was initially developed to secure provenance metadata, so clinical trial audit trails are a natural use case.  Scrybe provides the five basic properties of a provenance system listed in Section~\ref{sec:background}. The framework we propose here uses an uncoupled inversion-based changelog secured by Scrybe's annotation-based provenance framework. 
	
	This section will describe the data structures used in Scrybe and its mining method, keeping in mind the required properties. Scrybe uses a blockchain since the immutability property of blockchains provides strong integrity guarantees. The replication of the blockchain state among all miners also provides strong availability guarantees. The challenge is to design a system that maintains the other properties. A publicly visible blockchain makes access control difficult, and PoW mining is extremely inefficient. Figure~\ref{fig:case2} shows the Scrybe architecture.
	
	The main component of this architecture is the Scrybe blockchain and the changelog. We store the history of all changes made to institutional database records as entries in a changelog server.  A firewall can be placed around the changelog, and only qualified people may access it, ensuring HIPAA compliance. Since this server is not publicly accessible, patient privacy is protected. As long as the blockchain remains immutable, the changelog's integrity is guaranteed, and the changelog is a secure audit trail tracking every event in the clinical trial.
	
	The changelog entry cryptographic hashes are stored in the blockchain to ensure that data integrity and non-repudiation are guaranteed despite a centralized changelog server. Because of the immutability property of the blockchain, these hashes cannot be altered. When an audit is performed, the changelog can be examined, and the changelog hashes can be compared to the hashes stored on the blockchain. If there is a mismatch, then the auditor knows tampering has occurred. In the case of an FDA investigation, the blockchain transactions matching changelog entries that describe each record modification guarantee that the changelog is a trustworthy audit trail.  
	
	Each \textit{changelog entry} is recorded on the blockchain as a \textit{transaction}, and \textit{transactions} are grouped in \textit{blocks.} These blocks constitute the blockchain data structure.
	
	\subsection{Changelog Entry}
	A changelog entry describes a single change made to the secure REDCap database. These entries are not a part of Scrybe--they are an entirely separate primitive used to extend REDCap (or any secure database) with additional provenance functionality. Whenever a user performs any addition, deletion, or modification, an entry is created and stored in a  changelog server. The changelog server is a sequence of these entries. Each entry has an associated ID that increases sequentially with each new entry added to the log. The entry contains a modification field describing the change made to the database. The entry is signed to guarantee non-repudiation. Changelog entries can be applied sequentially (up to the most recent changelog entry) to an empty database to construct the current database. The changelog is stored locally by the institution and is not a part of Scrybe. Keeping the changelog server behind an institutional firewall alongside the REDCap database used for the clinical trial will ensure HIPAA compliance since only authorized personnel can view the data. When a changelog event occurs, the changelog server submits a transaction to the Scrybe blockchain containing a hash of the changelog entry and non-identifying metadata, such as date, time, and trial ID.
	
	\subsection{Transactions}
	Scrybe transactions contain whole pieces of provenance metadata. Any correlations to other pieces of metadata must be performed at a higher level outside the system. In the case of clinical trials that are performed with an institutional database, the changelog entries are secured by transactions.  Whenever a change is made to a record, an entry is stored in the changelog server. Then, a transaction with the changelog hash is broadcast to the blockchain miners.  This transaction contains a hash of the created entry, the time, and the entry's ID. The hash is used instead of the actual entry to ensure HIPAA compliance.
	
	When a transaction is created, it is cryptographically signed by its creator, the changelog server. The signature guarantees that if a transaction exists with a particular timestamp, entry hash, and valid signature, the signatory modified the database in the manner described by that particular entry. The transaction is the foundational building block of secure provenance. 
	
	\subsection{Blocks}
	A block contains a group of transactions, the hash of the previous block, and a  record of the mining process. Since each block contains a previous block's hash, an attacker must modify all previous blocks to modify the current block. Since the probability of producing a forged block is \textit{almost surely zero} (i.e., probability of 0), the blockchain can be considered immutable. The block size is determined by the volume of transactions and the time required to \textit{mine} a block. The time between blocks is adjusted to maintain a reasonable interval for transactions to accumulate. Each block contains a signature of the miner that generated the block.  The signature provides an added layer of security for transactions on the blockchain--the block itself contains a signed timestamp, and each transaction also contains a signed timestamp. Both of these attributes provide interleaved trust, allowing us to trust the timestamp on the changelog entry. Authorized participants can view the transaction, but no one can change it or deny its existence.
	
	
	\subsection{Mining}
	Miners generate the blocks in a blockchain and broadcast them to the rest of the miners.  Traditionally, PoW mining is too resource-intensive, requiring the constant generation of hashes until the miner solves a cryptographic puzzle. Proof-of-Stake (PoS) systems require that miners \textit{stake} a sizable amount of currency that will be forfeit if malfeasance is found.  While PoS scales better and provides a more economical approach, it requires a native currency and poses centralization risks if a particular miner controls a majority of the currency. 
	
	Scrybe is a permissioned blockchain--only authorized miners can generate blocks. Each block is signed, and any block signed by an unauthorized miner is immediately discarded. This lightweight hybrid PoS mining approach allows miners to \textit{stake} their participation in the mining process \cite{brooks2018scrybe}. A miner is randomly selected from the pool of authorized miners, and that miner is delegated to produce the next block. In practice, miners consist of various companies, research institutions, and regulatory agencies. A detailed description and rigorous proof of Scrybe's consensus algorithm is given in \cite{worley2020scrybe,scrybeproof}. Since Scrybe is a dedicated provenance blockchain, no underlying currency can introduce volatility or cause the underlying technology to become obsolete. Scrybe is not based on PoW or PoS, so there are no concerns with environmental impact or \textit{greedy nodes}. No leader or initial group of nodes is elected like in PBFT-based algorithms, ensuring the process is truly decentralized. Scrybe leverages the advantages of a permissioned blockchain with a secure and scalable algorithm. The complexity is $O(n)$ as the number of nodes approaches infinity \cite{scrybeproof}. 
	
	\section{REDCap Secured with Scrybe}  \label{sec:clinicaltrial}
	A proof-of-concept prototype was created to test the ideas behind the Scrybe framework. The core of the prototype is  Scrybe, which was implemented in C++. All data structures are stored as serialized strings in a cached database for fast access, and peer-to-peer communication happens asynchronously. We used REDCap as our institutional database for storing clinical trial data. The changelog server that secures REDCap communicates with a Scrybe client to submit transactions. The changelog server also includes tools for browsing the changelog database and verifying the integrity of the entries it contains by comparing their hashes with those stored on the blockchain. 
	
	For this application, an interface was created to input data into the system. REDCap exposes an API of HTTP POST requests that allows for data import and export \cite{api}. A Python script was written that allows clients to input data. That data is then simultaneously imported in REDCap through its API, uploaded to the changelog, and transactions securing the entries are submitted to the Scrybe miners. This interface is currently a standalone command-line interface, but REDCap has a tool for creating data entry interfaces.  Future work should consider a daemon that monitors the REDCap database and automatically creates changelog entries whenever changes occur. Automatic changelog generation would make the provenance backend invisible to the end-user, allowing for seamless integration.
	
	We present the use-cases shown below to illustrate the properties outlined in Table~\ref{tab:cfr}. To satisfy \cite{regulations}, we must address integrity, authentication, availability, access control, and non-repudiation. These use-cases introduce several new concepts. The Scrybe provenance consortium is a group of independent institutions using the Scrybe provenance framework to secure their respective data. The consortium is most robust when the independent institutions each have a Scrybe node and are unlikely to collude. The auditor represents any authorized individual who wishes to validate data. The researcher is an individual authorized to store data in the Scrybe provenance framework. In these use cases, the attacker is an unauthorized individual who is making malicious changes. These use-cases require public-key cryptography, and there are existing solutions and best practices for enterprise key management.
	
	\subsection{Integrity}
	
\begin{figure*}
		\centering
		\includegraphics[scale=1]{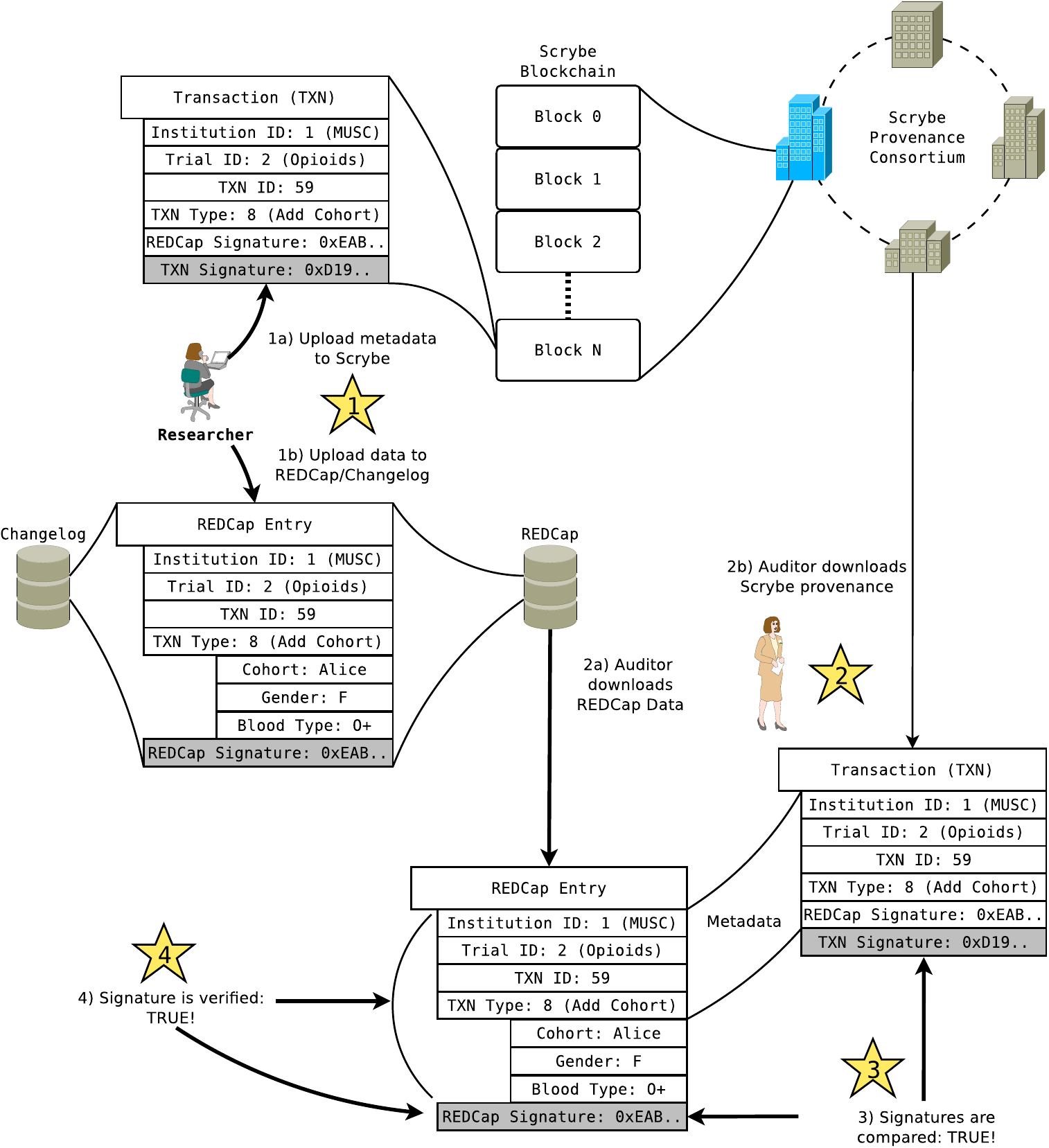}
		\caption{Verifying valid REDCap data using the Scrybe provenance framework.}
		\label{fig:case2}
	\end{figure*}

	Consider the standard use-case, shown in Figure~\ref{fig:case2}. A researcher uploads data to REDCap using the Scrybe provenance framework. The Scrybe provenance framework breaks the data into a series of changes that can be incrementally applied to the REDCap database. The metadata for each change is used to create a transaction, and this transaction is signed using the researcher's public key (TXN Signature). Simultaneously, each incremental change is also signed using the researcher's private key (REDCap Signature). The Scrybe framework stores the signed changes in a changelog\footnote{The changelog is stored on a secure server operated by the institution.}, allowing the current state of the database to be reconstructed.  Then, the Scrybe transaction is submitted to the miners, where it is incorporated into the blockchain. Finally, the data is uploaded to REDCap.
	
	When an auditor verifies the data stored in REDCap, the first step is to download a copy of the REDCap data and the corresponding Scrybe transactions. The transaction signature is used to verify the integrity of the metadata in the transaction. Next, the REDCap signature is compared to the REDCap signature stored in the Scrybe transaction. Once this signature is verified, it is used to verify the integrity of the REDCap entry. With this verification complete, the auditor can reconstruct the complete history using the changelog. Auditors may elect to use software tools to identify anomalies, such as conflicting changes, which may indicate intentional malfeasance on the researcher's part.
	
\begin{figure*}
		\centering
		\includegraphics[scale=1]{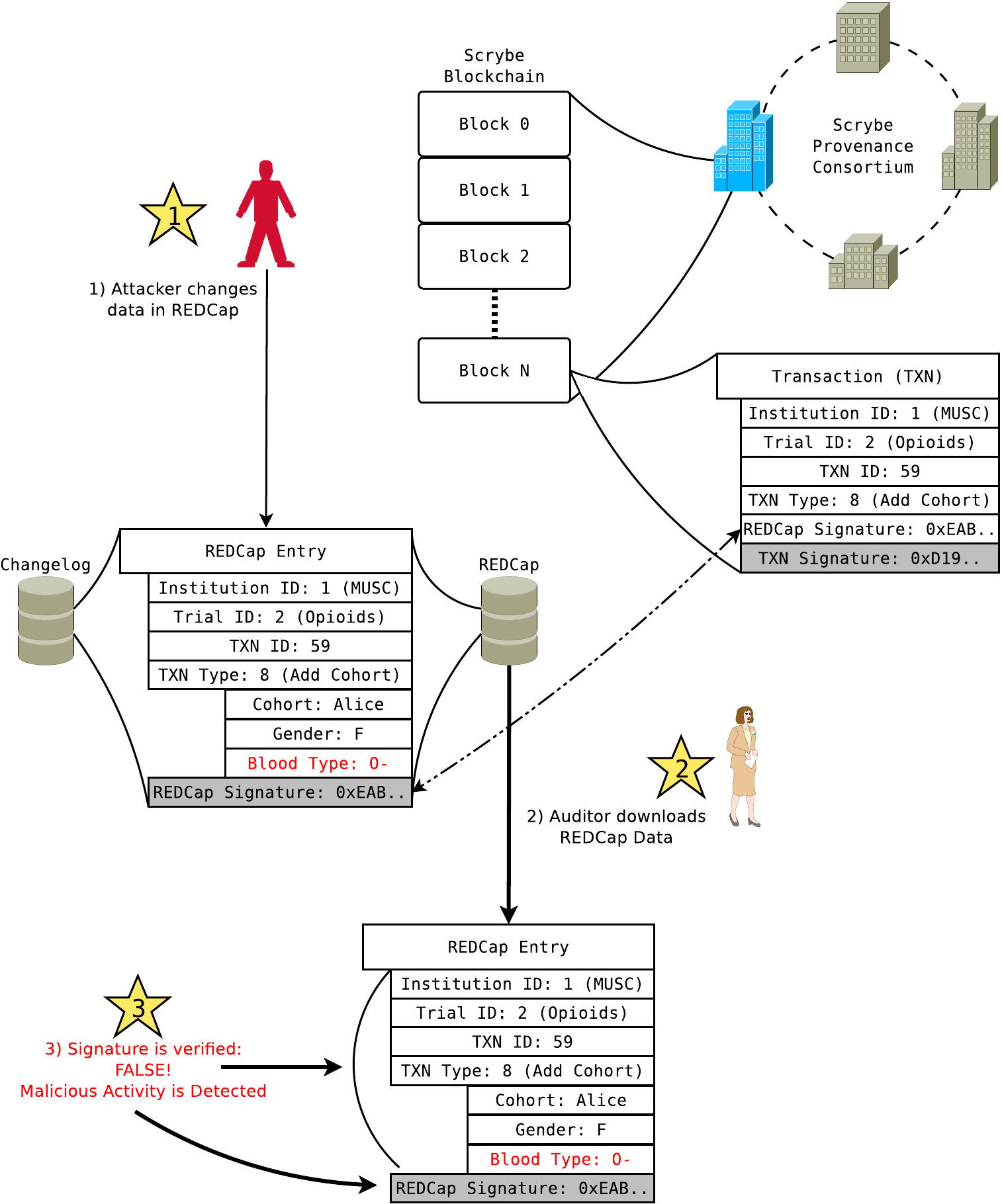}
		\caption{Detecting altered REDCap data using the Scrybe provenance framework.}
		\label{fig:case1}
	\end{figure*}

	Consider a malicious user modifying data in the REDCap database without the proper approval. This is shown in Figure~\ref{fig:case1}. The attacker modifies the blood type of a patient. There are two cases, 1) the attacker does not attempt to forge the REDCap signature, and 2) the attacker is a researcher and maliciously updates the database to include false information using the Scrybe framework. The latter case can be addressed by software that scans the changelog and identifies conflicting anomalous behavior. In the former case (shown in Figure~\ref{fig:case2}, the auditor downloads a copy of the REDCap data and verifies the signature. The signature is calculated using all of the fields stored in REDCap.  Section~\ref{sec:integrity} provided a proof showing that if any of the data changes, it is detected with the signature. In this scenario, the auditor cannot verify the REDCap signature, showing the data in REDCap was altered.
	
	The final scenario addresses malicious changes made to the Scrybe blockchain. The prerequisite is that an attacker compromises every Scrybe node. This compromise would include Scrybe nodes hosted at private, public, and federal institutions (e.g., the FDA, DHEC, and CDC). By compromising these institutions, the attacker gains access to the private keys the Scrybe nodes use to sign blocks. The attacker must also compromise the researcher who signed the transaction they wish to modify, allowing the attacker to forge the REDCap signature and the transaction signature. With the forged transaction signature, the attacker can reconstruct the entire blockchain using forged signatures. To our knowledge, there are no security countermeasures that can handle all of the nodes being compromised.
	
	
	\begin{figure*}
		\centering
		\includegraphics[scale=1]{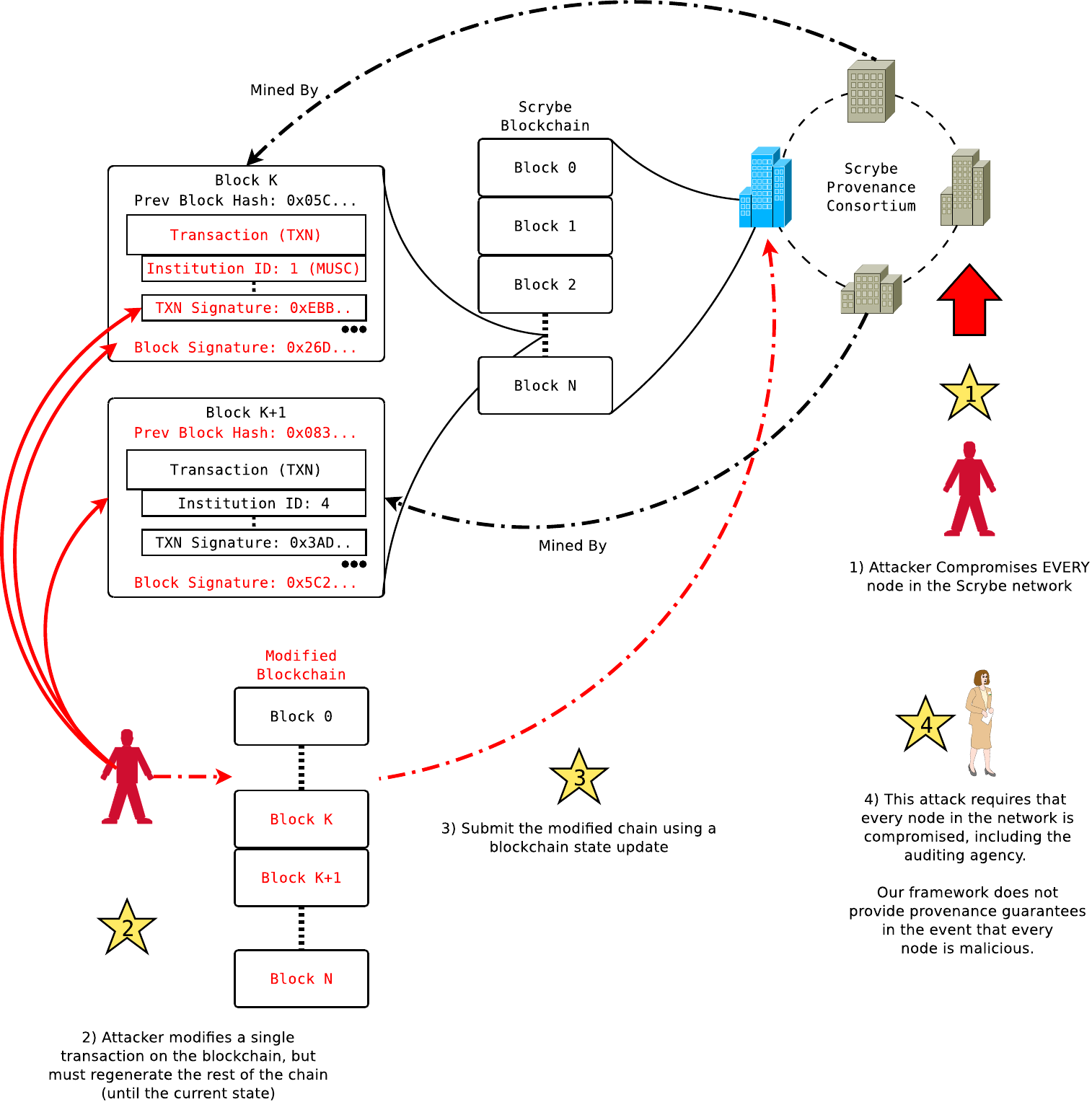}
		\caption{Detecting an total system compromise.}
		\label{fig:case3}
	\end{figure*}
	
	
	The general provenance requirements were distilled from Table~\ref{tab:cfr}. Scrybe satisfies each of these requirements in order to provide a secure framework for data provenance.
	
	\subsection{Availability}
	There are two aspects of availability: the original data's availability and the audit trail's availability. Scrybe only stores the audit trail, and storing the actual clinical trial data is outside the scope of Scrybe.  Since Scrybe is a distributed ledger, there are redundant copies stored at various sites. The Scrybe use-case recommends that each institution and regulatory oversight entity host a Scrybe node. Each of these nodes contains a copy of the blockchain. Further, it is recommended that the Scrybe miners be hosted on a server with modern storage redundancy features, such as RAID, which allows data to be recovered in the event of multiple hard drive failures. With these considerations, the bare minimum availability provided by Scrybe matches the current level of availability. However, these features ensure that a valid version of the provenance proof will be available in almost every failure or tampering event.
	
	\subsection{Authentication and Access Control}
	Scrybe is a distributed blockchain and has no central database to control. Instead, the distributed mining process determines what information is added to the blockchain. There is no need to assign privileges to modify on-chain data. Instead, each researcher has a public-private keypair that is publicly registered with the blockchain. Each researcher can only submit transactions when they are signed with a valid key, and these transactions are incorporated into the blockchain by a miner with a valid key. This signature ensures that only authorized researchers create transactions on the blockchain.
	
	Assigning rights to modify off-chain data is handled by the institutional database itself.  This access control is not affected by our approach and can be done using existing methods. Care must be taken to ensure that all off-chain data modification is linked with an on-chain record.  Since only hashes of the changelog entry are stored on the blockchain, there is no need to restrict read access to the transactions. In addition to simplifying the system, this guarantees HIPAA compliance.
	
	\subsection{Efficiency}
	Section~\ref{sec:related} discusses the several approaches solutions in this space employ. Most popular blockchains, such as Ethereum and Bitcoin, use a PoW mining algorithm. Even permissioned Ethereum blockchains still currently use PoW. Many provenance blockchains leverage existing technology, such as Bitcoin and Ethereum. These mining algorithms are designed to consume all available resources wherever it is running.  Using Bitcoin or Etherereum as the basis of a provenance blockchain is economically inefficient and morally irresponsible. Despite this limitation, there are still issues with scalability and volatility. Other solutions, such as Hyperledger's built-in consensus algorithms, face challenges with scalability and decentralization. There are also a set of solutions that employ novel blockchain solutions. As discussed in Section~\ref{sec:related}, many of these solutions have scalability and security issues.
	
	Scrybe uses a permissioned blockchain that is built with a novel lightweight mining algorithm \cite{brooks2018scrybe}. Each miner only expends energy communicating and mining when selected through a non-deterministic algorithm \cite{brooks2018scrybe}. By exchanging signed messages containing all the received transactions, nodes can validate the block published by the selected node. As part of the message exchange, a new node is chosen to produce the next block. The algorithm is described in detail in \cite{worley2020scrybe,scrybeproof}. This algorithm was proven to have complexity $O(n)$ as $n$ approaches infinity \cite{scrybeproof}. Scrybe is an efficient alternative consensus algorithm that is capable of registering provenance for multiple clinical trials.
	
	
	\subsection{Clinical Trial Data Validation}
\begin{figure*}
		\centering
		\includegraphics[scale=0.3]{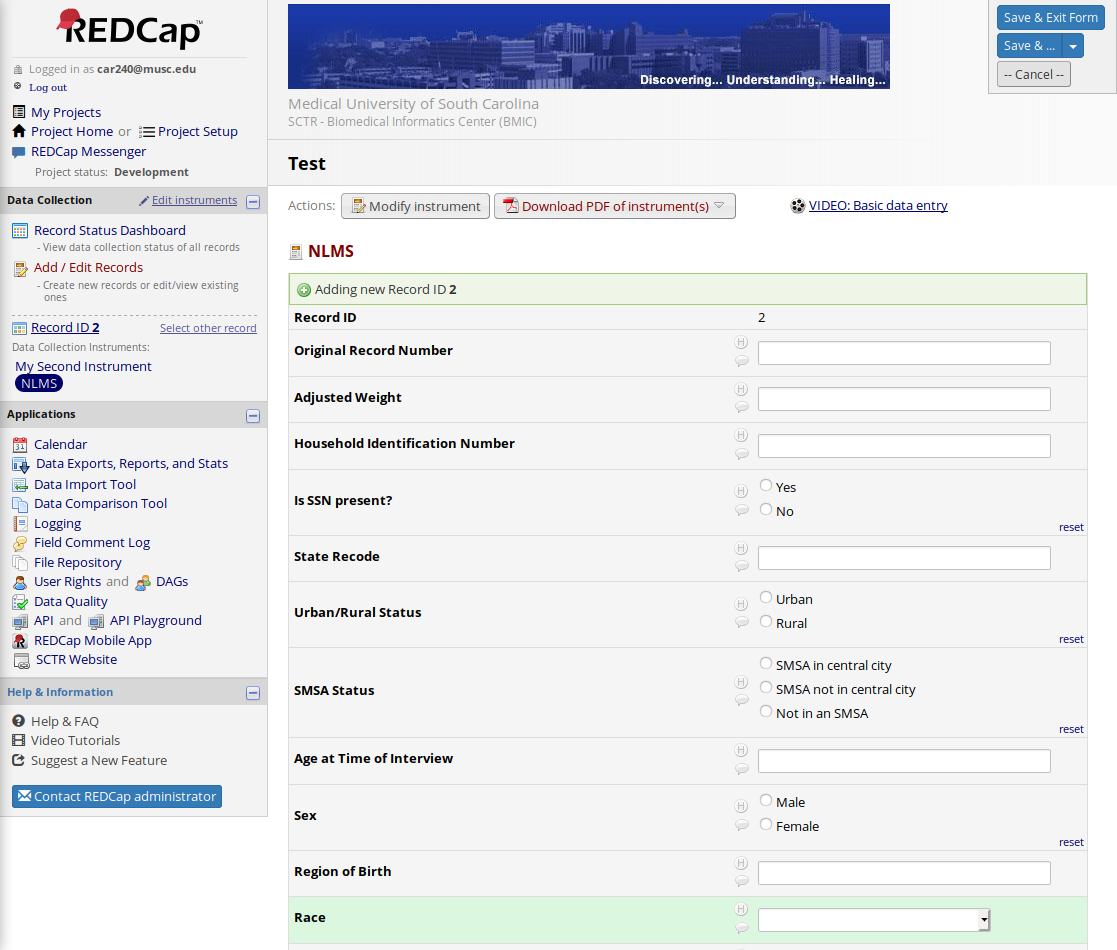}
		\caption{REDCap data input instrument for the NLMS dataset.}
		\label{fig:redcap_web}
	\end{figure*}

	For experimental purposes, the public-use National Longitudinal Mortality Study (NLMS) dataset was acquired from the National Institute of Health's Biologic Specimen and Data Repository Information Coordinating Center \cite{dataset}. This large-scale dataset relates mortality to many lifestyle factors, such as age, location, or substance use. A REDCap instance at the Medical University of South Carolina was used to store the data. REDCap instruments were created for this dataset, allowing the data to be input and processed. An example REDCap input interface is shown in Figure \ref{fig:redcap_web}. 
	
	The interface script was used to submit the NLMS data in CSV format to REDCap and the changelog server. The data can be viewed by exporting individual entries in the Scrybe command-line interface. The changelog server's pull option can be used to download the entire changelog and the entire blockchain to conduct an item-by-item comparison to audit integrity. If there is an entry in the changelog with no corresponding transaction in Scrybe, a warning is given. An error is raised if an entry's value does not match the hash in the corresponding transaction. Similarly, if a transaction is modified on the local blockchain database, anyone can verify that the transaction is invalid due to the invalid "previous block" hash in the following block. If a recent transaction or set of transactions is missing, synchronization with the rest of the miners resolves the issue.
	
	\begin{figure}
		\centering
		\includegraphics[scale=0.45]{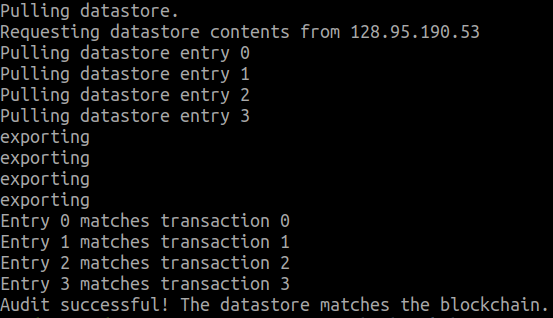}
		\caption{Successful comparison of changelog (datastore) and blockchain.}
		\label{fig:pull1}
	\end{figure}
	
\begin{figure}
		\centering
		\includegraphics[scale=0.45]{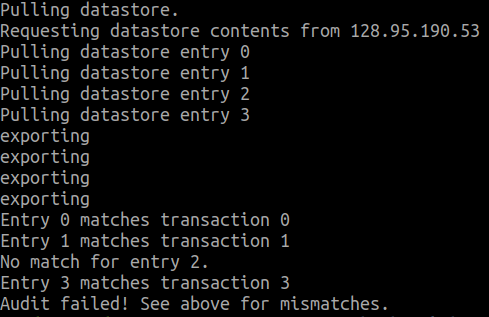}
		\caption{On a corrupted local instance, the audit failed.}
		\label{fig:pull2}
	\end{figure}

	Figure~\ref{fig:pull1} shows an example of a successful comparison between the blockchain and the changelog. Every changelog entry hash was stored in a block transaction.  The blockchain was manually corrupted, so the hash of an entry would no longer match a valid changelog entry. This corruption was identified, as shown in Figure~\ref{fig:pull2}, validating the integrity of the changelog.
	
	\section{Discussion} \label{sec:discussion} 
	Part 11 of Title 21 Code of Federal Regulations \cite{regulations} and ISO 27789 \cite{iso27789} requires that researchers guarantee the authenticity, integrity, and confidentiality of data collected for clinical trials. Clinical trials are one of the most important forms of scientific research mechanisms for advancing human health, and the FDA closely regulates them in the United States of America. The increased use of smart and wearable IoT devices in clinical trials presents a unique challenge: the advent of computerized data management in EDC  and CDM systems has not yet been adequately addressed.
	
	We propose Scrybe, a permissioned blockchain, as a method of storing proof of data provenance. Scrybe uses a lightweight mining algorithm that is more efficient and economical than popular proof-of-work algorithms (e.g., Ethereum and Bitcoin). Many existing solutions based on popular cryptocurrencies are subject to additional overhead and volatility. These solutions are also tied to the cryptocurrency's future success (or failure).  Scrybe is more decentralized than consensus algorithms based on pBFT, which have become popular in the provenance blockchain space. By using a distributed consensus among competitors, Scrybe ensures immutability. Considering the requirements outlined in \cite{regulations}, we demonstrate how Scrybe addresses each of the relevant controls. A proof-of-concept integration with REDCap is used to show tamper resistance. The REDCap-Scrybe provenance framework allows researchers to track the provenance of any clinical trial data collected by smart devices. 
	
	
	
	Future work will include further integration with REDCap and trial runs on more datasets. The Scrybe transaction process will be integrated as a separate daemon that monitors the REDCap database, automatically generates changelog entries, and submits a transaction whenever changes are detected, providing seamless integration with existing EDC systems. As discussed in Section~\ref{sec:related}, Hyperledger offers pluggable consensus algorithms. Leveraging the Hyperleger framework and implementing a pluggable Scrybe consensus algorithm would leverage existing technology with a strong community. Scrybe's application is not limited to tracking clinical trial provenance. There are other projects currently leveraging this technology. A future version of Scrybe will include smart-contract functionality to provide researchers with additional functionality and provenance security.
	
	\section*{Acknowledgment}
	This project was supported, in part, by the National Center for Advancing Translational Sciences of the National Institutes of Health under Grant Number UL1 TR001450, the South Carolina SmartState Program, and National Science Foundation grants CNS-1049765, OAC-1547245, and CNS-1544910. The U.S. Government is authorized to reproduce and distribute reprints for Governmental purposes, notwithstanding any copyright notation thereon. The authors gratefully acknowledge this support and take responsibility for the contents of this report. The views and conclusions contained herein are those of the authors and should not be interpreted as necessarily representing the official policies or endorsements, either expressed or implied, of the National Institutes of Health, the National Science Foundation, or the U.S. Government. 
	
	\bibliographystyle{IEEEtran}
	\bibliography{bioinformatics}

\end{document}